\documentclass{elsart1p}
 \usepackage{epsfig}
 \usepackage{epstopdf}
\usepackage{graphicx}
\usepackage{amssymb}
\begin{document}
\begin{frontmatter}
% notes:
% use the corauthref command within \author for corresponding author
% footnotes;
\title{Two particle correlations with photon and $\pi^{0}$ triggers with ALICE}
\thanks{\small Supported by NSFC (10875051, 10635020 and 10975061), the Key Project of Chinese Ministry of Education (306022), the Program of Introducing Talents of Discipline to Universities of China (QLPL200909, B08033 and CCNU09C01002)}

\author{{\bf Yaxian Mao}$^{1,2}$, for the ALICE Collaboration}
\address{$^{1}$ Key Laboratory of Quark $\&$ Lepton Physics~(Huazhong Normal
University),  Ministry of Education, Wuhan 430079, China}
\address{$^{2}$ Laboratoire de Physique Subatomique et de Cosmologie, Grenoble 38026, France}
\begin{abstract}
%Hard scattered partons produced in initial stage of the collisions, have been identified as a valuable probe of the medium created in heavy-ion collisions. 
%The medium properties can be inferred from the modifications experienced by the partonic shower inside the medium. 
Comparing the measurements of the hadronic final state from partonic showers in proton-proton and heavy-ion collisions will reveal the modifications generated by the medium on partons produced in hard scatterings.
This can be achieved by selecting the hard processes in which there is a direct photon in the final state.
The experimental technique consists in tagging events with a well identified high energy direct photon and in measuring the azimuthal angle correlation with charged hadrons.  
To establish a reference measurement for heavy-ion collisions, proton-proton collision data collected with ALICE have been analyzed. Preliminary results are presented together with photon and $\pi^{0}$-charged hadrons correlations showing the characteristic di-jet pattern from where the partonic momentum $k_{T}$ is extracted. 
%Implications for the heavy-ion data analysis are also discussed. 
\end{abstract}
\begin{keyword}
Triggers, Azimuthal Correlation, Isolation Cut, $k_{T}$
\PACS 21.10.Hw, 25.75.Gz, 21.10.Hw, 12.38.Mh
\end{keyword}
\end{frontmatter}

% main text
\section{Introduction}
High energy heavy-ion collisions enable the study of strongly interacting matter
under extreme conditions. 
At sufficiently high collision energies Quantum-Chromodynamics (QCD) predicts that hot and dense deconfined matter, commonly referred to as the Quark-Gluon Plasma (QGP), is formed.
The experiment ALICE~\cite{ALICE} at the  CERN Large Hadron Collider (LHC)~\cite{LHC}, allows the study of the QCD matter in a new energy domain.

High $p_T$ partons produced in the initial stage of the collisions,  have been identified as a valuable probe of the medium. 
%Indeed, medium properties can be inferred from the modifications experienced by the partonic shower inside the medium. 
They are only observed indirectly, as a collimated jet of hadrons originating from the hadronization of the partonic shower.
Comparing the properties of the jet fragmentation in proton-proton and heavy-ion collisions will reveal the modifications induced by the medium on the hard scattered partons.
Ideally, one needs to know the 4-momentum of the parton when it has been produced in the hard scattering and after it has been modified by the medium.
This can be achieved by selecting particular hard processes in which there is a photon in the final state.
Since the photon does not interact with the medium, its 4-momentum is not modified and thus provides a measure of the hard scattered parton emitted back-to-back with the photon.
Measuring the hadrons opposite to the photon is thus a promising way to measure the jet fragmentation and imbalance between photon and hadrons to quantify the modifications due to the medium. 

To establish a reference measurement for heavy-ion collisions, proton-proton collision data collected with ALICE in 2010 have been analyzed with the ultimate goal to construct the direct photon-charged hadron correlations. Minimum bias data have been collected in pp collisions at center of mass $\sqrt{s}~=7~$TeV. The present results have been obtained by  analyzing about 160 million events. 
The preliminary result is presented together with inclusive photon-charged hadrons correlation and $\pi^{0}$-charged hadrons spectra all showing the characteristic di-jet pattern from where the momentum imbalance $k_{T}$ is extracted. 
\section{Trigger selection}
The experimental technique consists in tagging events with a leading trigger and measuring the distribution of hadrons associated to this leading trigger from the same event. 
Such a measurement requires an excellent
photon and $\pi^{0}$ identification and the measurement of charged and
neutral hadrons with good $p_{T}$ resolution. In ALICE, the
electromagnetic calorimeters, PHOS ($|\Delta\eta|<0.12$ and $\Delta
\phi$ =100$^o$) and EMCal ($|\Delta\eta|<0.7$ and $\Delta \phi$
=100$^o$)~\cite{TDR}, are capable to measure photons with high
efficiency and resolution. 
In the calorimeters, electromagnetic particles are detected as clusters of cells in the calorimeters. Roughly we have identified $\pi^{0}$ candidate as a pair of clusters with invariant mass around  the $\pi^{0}$ mass between 110 and 160~MeV/c$^2$, and single clusters are identified as inclusive photon candidates.
No particle identification has been applied yet so that the single cluster sample may contain a sizable fraction of charged particles which develop a shower in the calorimeters or high-$p_{T}$ merged $\pi^{0}$ cluster which can not be reconstructed by invariant mass.
The central tracking system (ITS and TPC), covering the pseudo-rapidity
$-0.9 \leq \eta \leq +0.9$ and the full azimuth, is used for charged track measurements, contributes to the direct photon identification by applying the isolation technique. 

Three different trigger particles have been selected for the correlation measurements: (i) the charged trigger is chosen as the track with highest transverse momentum among all the tracks from the same event, (ii)  the photon cluster trigger is defined as the calorimeter cluster with highest energy and no charged track from the same event has momentum larger than photon cluster, (iii)  the $\pi^{0}$ trigger is selected as the cluster pair within the appropriate invariant mass range and with the highest transverse momentum in the event. 

\section{Azimuthal Correlation}
\begin{figure}[h]
\hspace {1pc}
\begin{minipage}{15pc}
\includegraphics[width=15pc]{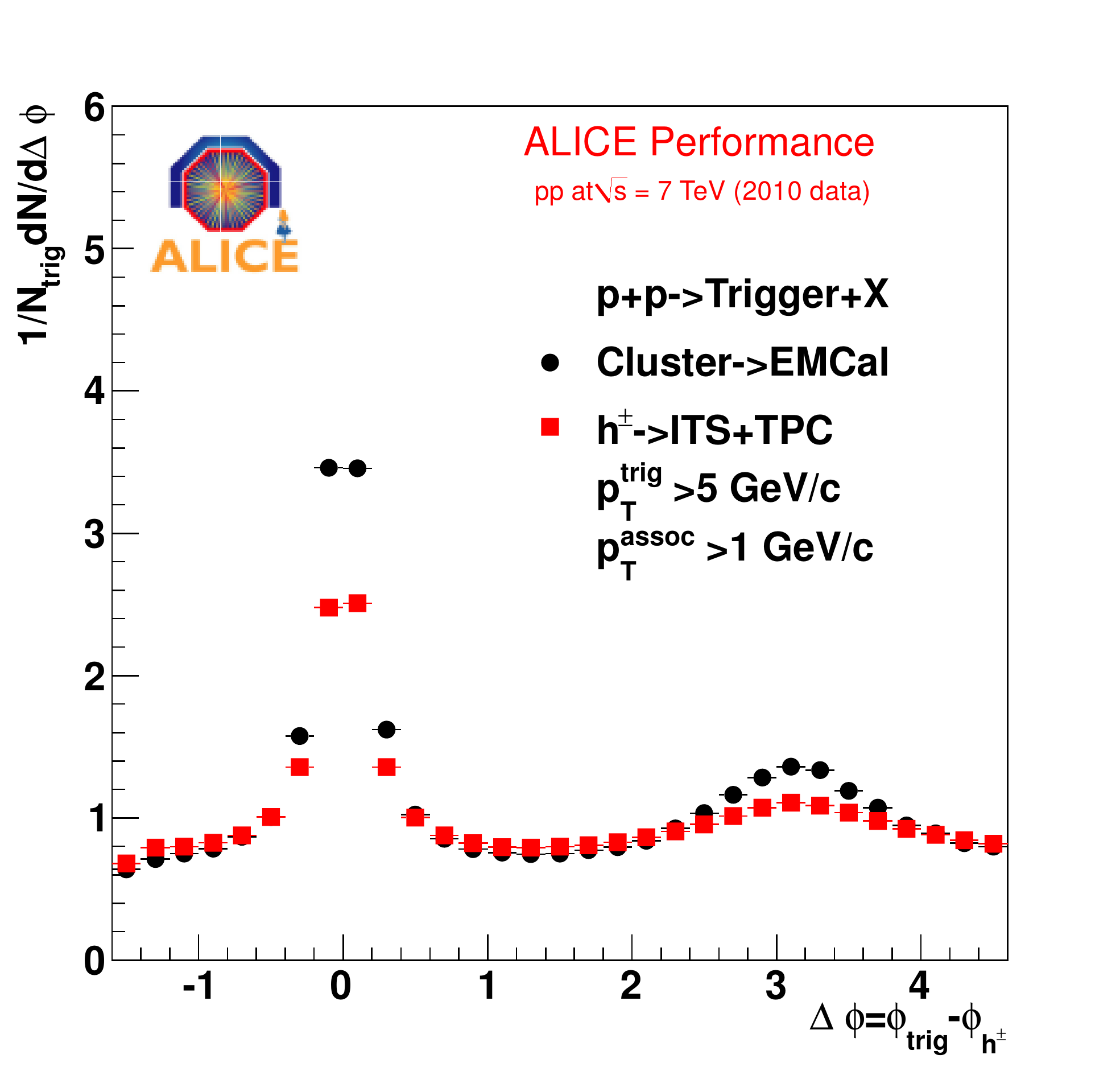}
\caption{\label{fig:ClusterCh}Relative azimuthal angle distribution $\Delta \phi = \phi_{trig}-\phi_{h^{\pm}}$ for charged trigger and inclusive cluster trigger with $p_{T}^{trig}  > 5$~GeV/c in pp collisions at $\sqrt{s}$~=~7~TeV. }
\end{minipage}\hspace{1pc}%
\begin{minipage}{15pc}
\includegraphics[width=15pc]{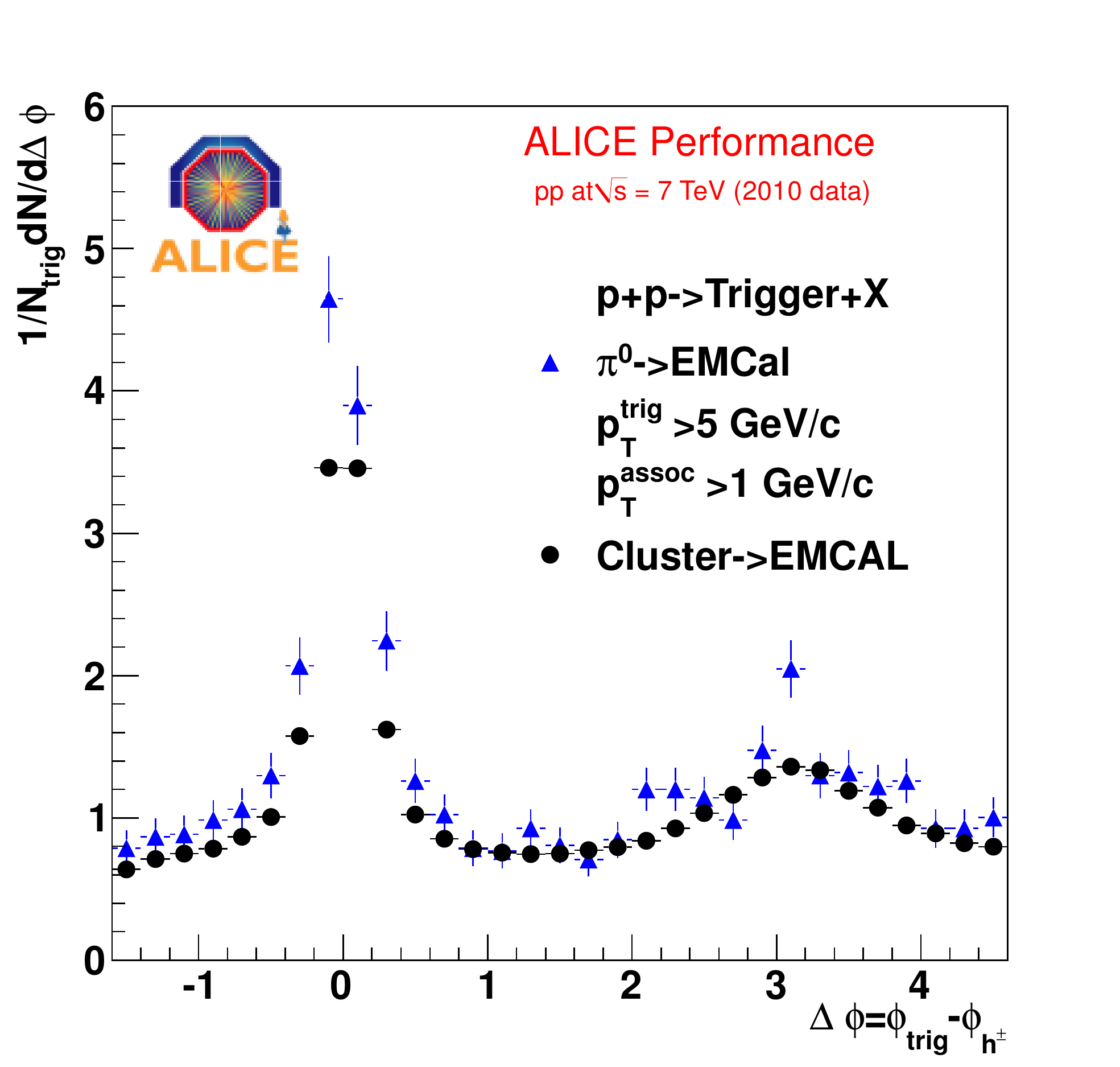}
\caption{\label{fig:PhotonPi0}Azimuthal correlation distributions for inclusive cluster trigger and $\pi^{0}$ triggers on the trigger particles with $p_{T}  > 5$~GeV/c in pp collisions at $\sqrt{s}$~=~7~TeV.}
\end{minipage}
\end{figure}

The azimuthal correlation between the trigger particles (charged particle, single cluster) and charged hadrons are shown in Fig.~\ref{fig:ClusterCh}. The near side ($\Delta\phi=0$) and away side ($\Delta \phi=\pi$) peaks are clearly observed. 
The correlation with cluster triggers shows larger  di-jet peaks reflecting the fact that the neutral trigger selection enhances the probability that the trigger is the leading particle of the jet fragmentation compared to the less restrictive charged trigger selection.
The azimuthal correlations from inclusive photon clusters and $\pi^{0}$ triggers show quite similar shapes (Fig.~\ref{fig:PhotonPi0}), indicating that most of the inclusive photon clusters are $\pi^0$ decay photons. 

By selecting isolated triggers, i.e. the trigger satisfies: the sum of the transverse momentum of the hadrons inside a cone with size $R = 0.4$ around the trigger candidate carries less than 10~\% of the trigger's transverse momentum, 
we can enrich the sample with direct photons or single hadron jets. 
The near side peak is suppressed by construction, whereas the away side peak remains and a slight difference is due to the imperfect isolation parameters used in the analysis (Fig.~\ref{fig:IsoEmcTrigDphi}) indicating the existence of di-jet events with one of the jet being a hard fragmenting jet or eventually a direct photon.  
 
%Note, however, that these distributions have not been corrected for efficiency. It is interesting to remark that at this very preliminary stage of the analysis we find that  the underlying event background level, outside the peaks region, is independent of the type of trigger particles, giving some confidence in the measurement. 

%However this preliminary analysis does not allow to %analysis is much too preliminary to draw any conclusion other that these results indicate the expected behaviour.
%The isolation parameters are not well adjusted and especially in our case only charged tracks are considered in our isolation cone due to the limited calorimeter acceptance (40~\% EMCAL and 60~\% PHOS have been installed so far).

\section{$k_{T}$ extraction}
\begin{figure}[h]
\hspace {1pc}
\begin{minipage}{15pc}
\includegraphics[width=15pc]{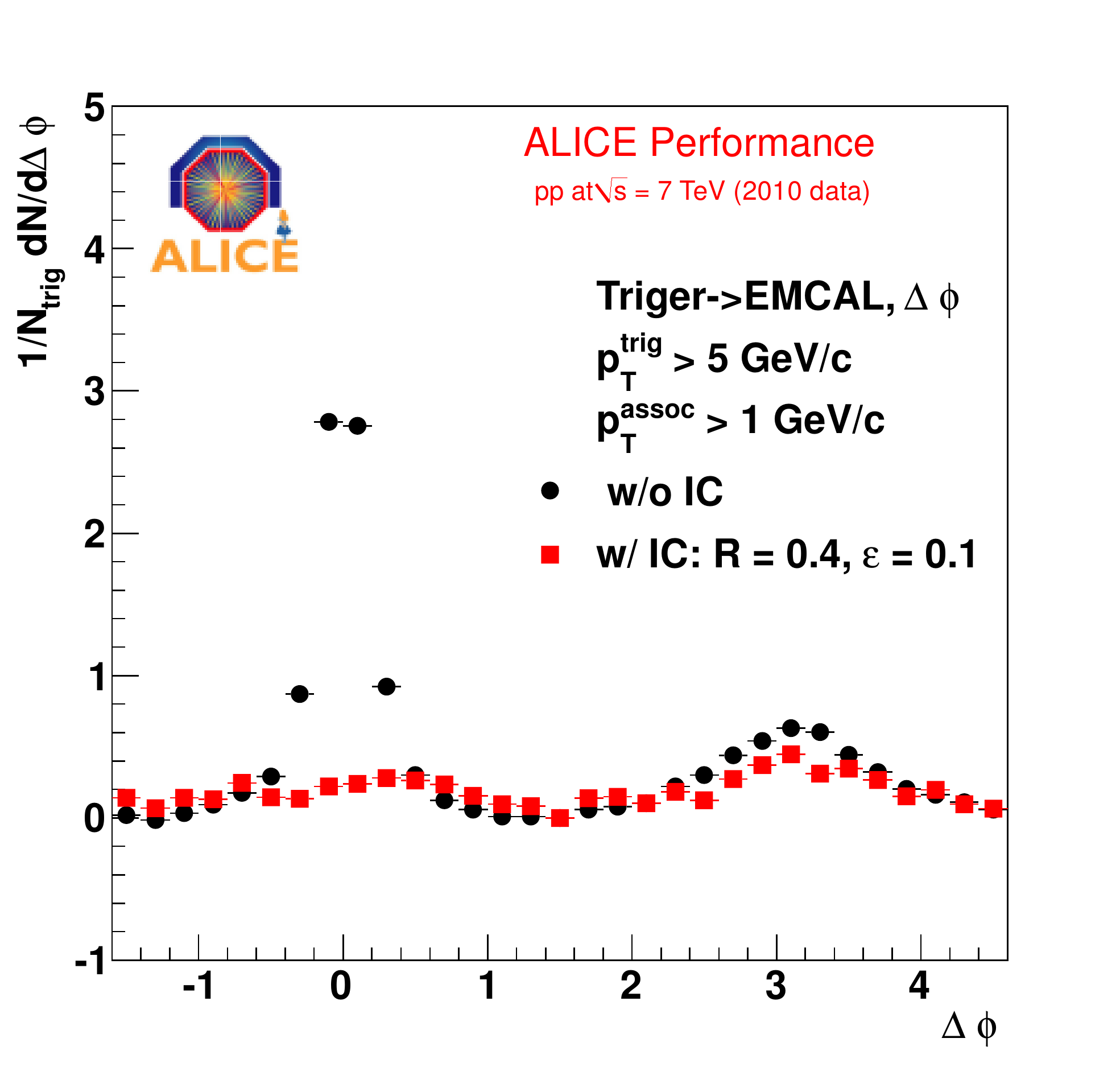}
\caption{\label{fig:IsoEmcTrigDphi}Azimuthal correlation distributions for inclusive cluster triggers with $p_{T}  > 5GeV/c$ before and after isolation cut (IC) selection: $R = 0.4$, $\varepsilon = 0.1$}
\end{minipage}\hspace{1pc}%
\begin{minipage}{15pc}
\includegraphics[width=16pc]{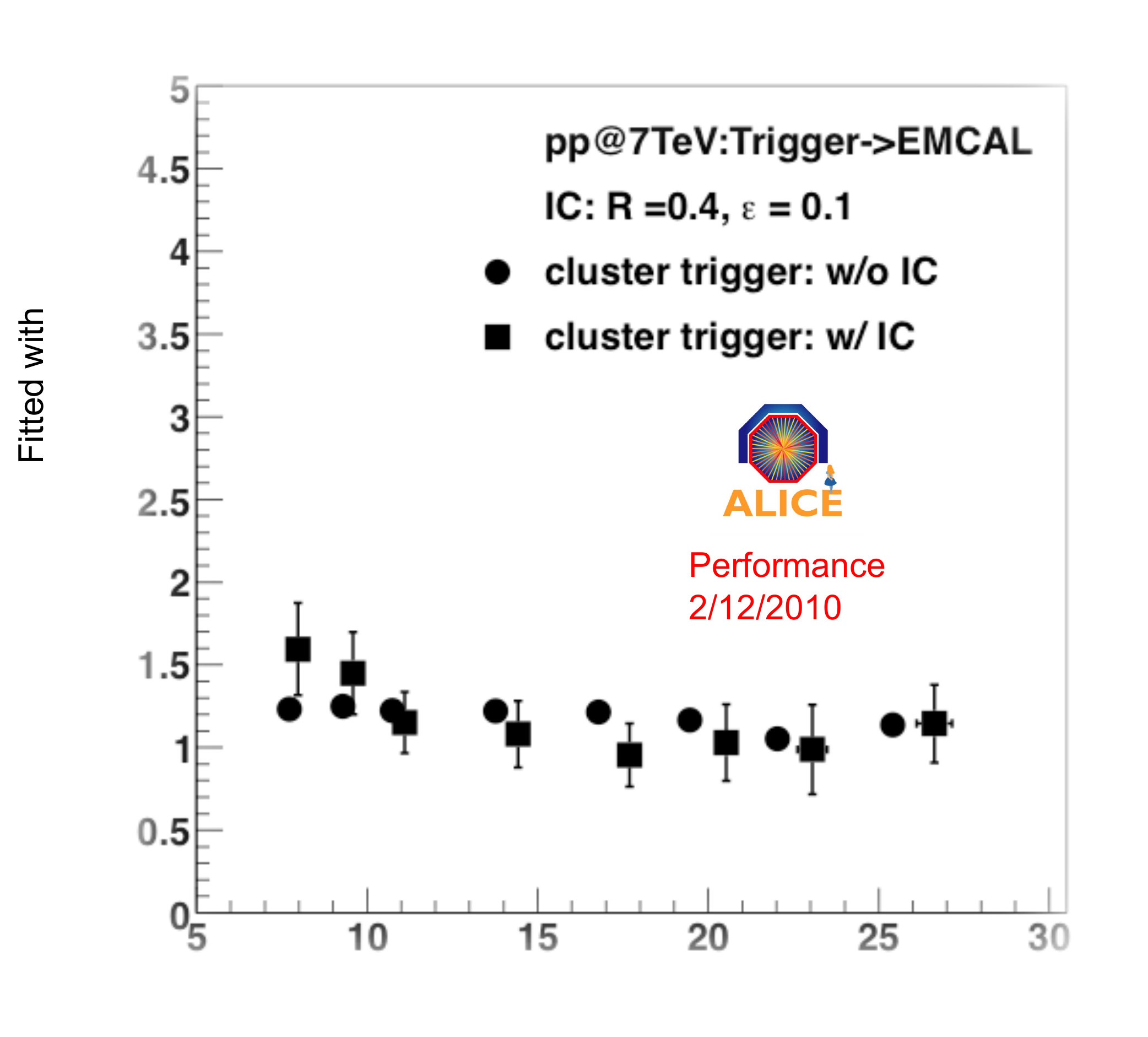}
\caption{\label{fig:pout}  Fitted width of the away side peak on the azimuthal correlation distribution with cluster triggers before and after isolation selection in EMCAL. }
\end{minipage}
\end{figure}

Because of the hadronization, we do not have direct access to the parton kinematics and therefore can measure neither the fragmentation function nor the magnitude of partonic transverse momentum $k_{T}$ which modifies the ideal $2 \rightarrow 2$ kinematics. 
However, the isolated photon/$\pi^{0}$ triggered correlation could be used to extract the partonic level kinematics to the extend that the Leading Order kinematics dominates, as suggested by the PHENIX analysis~\cite{kT}:
\begin{equation}
\frac{<z_t>}{\hat{x}_h}\sqrt{<k_{T}^{2}>} = \frac{1}{x_h}\sqrt{<p_{out}^{2}> - <j_{T_{y}}^{2}>(1+x_h^2)}\; 
\label{eq:kt}
\end{equation} 
where $z_t=\frac{p_T^{trig}}{\hat{p}_T^{trig}}$ is the trigger fragmentation variable  and $\hat{x}_h = \frac{\hat{p}_T^{assoc}}{\hat{p}_T^{trig}}$ is the ratio between away and near side hard scattered partons, $x_h$ is similar to $\hat{x}_h$ but at the hadronic level, and $j_{T_{y}}$ is the projection of trigger particle deviates from the parton before fragmentation (see detail in~\cite{kT}).  
The values of $\sqrt{<k_{T}^{2}>}$ are determined by measuring the width of the away side peak $\sqrt{<p_{out}^{2}>}$, using the fitting function described in~\cite{PHENIXCorr}. The fitted away side peak width shows in Fig.~\ref{fig:pout}, the width is weakly depend on the trigger $p_T$. 
The isolated trigger represents the hard scattered parton direction approximately ($\hat{p}_{T}^{trig}~\simeq~p_{T}^{trig}$), therefore, $z_{t}~\approx$~1 and $j_{T_{y}}~\approx~$0.  The $\sqrt{<k_{T}^{2}>}$ values we have measured for $p_{T}^{trig}  > 5~$GeV/c and $p_{T}^{assoc}  > 1~$GeV/c is consistent with another measurement obtained from charged di-hardon correlations in ALICE. The measured value agrees the extrapolated value at LHC energies from available worldwide data~\cite{QM2009}.    
 
However this preliminary analysis does not allow to draw any conclusion other than these results indicate the expected behavior.
Exciting physics will certainly come with the final analysis of large statistics within
well-calibrated detectors and all efficiency corrections.

\end{document}